\documentclass{svproc}
\usepackage{graphicx}%
\usepackage{multirow}%
\usepackage{amsmath,amssymb,amsfonts}%
\usepackage{mathrsfs}%
\usepackage[title]{appendix}%
\usepackage{xcolor}%
\usepackage{textcomp}%
\usepackage{manyfoot}%
\usepackage{booktabs}%
\usepackage{algorithm}%
\usepackage{algorithmicx}%
\usepackage{algpseudocode}%
\usepackage{listings}%
\usepackage{subcaption}
\usepackage{multicol}
\usepackage{adjustbox}
\usepackage{colortbl}
\usepackage{comment}
\usepackage{footmisc}
\usepackage{url}

\definecolor{darkeryellow}{cmyk}{0, 0.3, 1, 0}

\begin{document}
\mainmatter              
\title{Dynamics of Fisheries in the Azores Islands: A Network Analysis Approach}
\titlerunning{Azores Fisheries}  
%
\author{Brenda Nogueira\inst{1}\textsuperscript{*} \and Ana Torres\inst{1}\thanks{The authors contributed equally to this work.} \and
Nuno Moniz\inst{2} \and Gui M. Menezes\inst{3}}
\authorrunning{Brenda Nogueira, Ana Torres, Nuno Moniz, Gui M. Menezes} 
\institute{University of Porto, Faculty of Sciences, \\ Porto, Portugal \\
\and
Lucy Family Institute for Data \& Society, University of Notre Dame \\
United States
\and
Institute of Marine Sciences - OKEANOS,
University of the Azores,\\ Horta, Azores, Portugal}
\tocauthor{author list}

\maketitle              

\begin{abstract}
In the context of the global seafood industry, the Azores archipelago (Portugal) plays a pivotal role due to its vast maritime domain. This study employs complex network analysis techniques to investigate the dynamics of Azores fisheries, using time series data converted into networks. We uncover associations between Tunas and specific islands, consistent links among fish classifications, and identify other pivotal nodes within the fishing network. Remarkably, nodes with high degrees and a local clustering coefficient of one provide crucial insights into the fishing ecosystem. This study highlights the value of network analysis for understanding fisheries complexities and offers insights into sustainable management and the preservation of marine ecosystems. It also emphasizes the urgency for ongoing research and data collection to enrich our understanding of this multifaceted domain.
\end{abstract}

\section{Introduction}\label{introduction}
The demand for seafood has experienced a significant surge, with approximately 179 million tons of fish produced worldwide in 2018. Out of this, 156 million tons were used for human consumption, representing an annual supply of 20.5 kg per capita~\cite{fao}. 


The Autonomous Region of Azores (Portugal), consisting of nine islands spread over 600 km, significantly contributes to the size of Portugal's EEZ~\cite{sea}. Despite the expansive maritime territory, the waters around the Azores pose fishing challenges due to depth, currents, and seabed nature. For this reason, local fishing occurs near islands, banks, and seamounts less than 1,000 meters deep~\cite{LotacorSA}, resulting in more artisanal, multi-segmented fleets, that use varied gear, and a diversity of targeted species. Furthermore, annual landings average 11,000 tons (worth €33 million)~\cite{sea}, indicating the importance of fishing for local communities and the Azores' economy.

For these reasons, analyzing fisheries data can help understanding the complex marine interactions of the area, by exploring, not only temporal dynamics, but also visible patterns in different contributors, with the goal of aiding the decision-making process regarding fisheries practices.

However, it's crucial to note that this data presents an inherent complexity, extending into univariate and high-dimensional time series analysis, which continue to grapple with limitations across diverse contexts~\cite{Silva2021}. To address this, a promising solution emerges in the form of mapping time series onto networks, as these hold the potential to encapsulate intricate dependencies among constituent processes, encompassing both immediate and delayed inter-plays, as well as serial dependencies~\cite{Silva2021}. Therefore, this paper chooses to apply the methodology of transforming the time series of observations into temporal networks, and then to use tools of network analysis to uncover interesting insights.

This paper is structured as follows: Section 2 provides a background of fundamental concepts. Section 3 introduces the data and outlines the data preparation process for generating time series. Section 4 describes the methodology chosen to construct the networks and presents the results and discussion of the analysis, and, finally, in Section 5, the study's findings are summarized and discussed.

\section{Background Concepts}\label{literature}

A network or graph, represented as $G(V,E)$, is an ordered pair where $V$ represents the set of nodes (or vertices) and $E$ the set of edges (or links) between pairs of nodes belonging to $V$. A graph is most commonly represented as an adjacency matrix, denoted as $A$, in which an entry $A_{i,j}$ equals $1$, if there is an edge connecting the two nodes $i$ and $j$, or 0 otherwise~\cite{Silva2021}.

There are various approaches to construct a network from a time series or a set of time series. We will be focusing on using a set of time series for this transformation, which works by mapping states of the time series into nodes of the network and creating links between those nodes based on a measure of distance or similarity~\cite{Silva2021}. This process begins with the computation of the distance between all pairs of time series, resulting in a distance matrix ($D$).

In this case, we will be using Dynamic Time Warping (DTW) as the distance function, which aligns time series using a warping path, distinct from lock-step measures like Euclidean distance~\cite{Berndt1994UsingDT}. It optimizes the warping path to minimize the global warping cost, calculated through dynamic programming and a cumulative distance formula for two time series $X$ and $Y$ (both of length $T$), which is defined by:

\[\resizebox{\columnwidth}{!}{
    $d_{dtw}(X,Y)=dtw(i=T,j=T) = 
\begin{cases}
    \infty & \text{if } i=0 \oplus  j=0\\
    0              & \text{if } i=j=0\\
    \|X_i-Y_i\|+min
    \begin{cases}
    dtw(i-1, j)\\
    dtw(i, j-1)\\
    dtw(i-1, j-1)
    \end{cases} & \text{otherwise}
\end{cases}$}
\]

DTW stands out as a powerful technique for analyzing time series datasets due to its adaptability to variations and dynamic patterns, making it superior to rigid similarity measures. Its primary advantage lies in its invariance against shifting and scaling along the time axis. This unique feature has made DTW highly favored in pattern matching tasks. Notably, DTW not only provides a distance measure between two sequences but also offers insights into how these sequences are aligned with each other. In certain cases, understanding the alignment can be as informative, if not more so, than the distance itself ~\cite{cai2019dtwnet}.

Following the computation of these distances, the next step involves converting the distance matrix $D$ into an adjacency matrix $A$, that will represent the network. For this conversion, we can choose from a variety of methods:
\begin{itemize}
    \item \textbf{k-Nearest Neighbors Network (k-NN):} each node is connected to the $k$ other nodes with the shortest distances. This requires finding the $k$ closest elements for each row $i$ in $D$~\cite{ferreira22}.
    \item \textbf{$\epsilon$-Nearest Neighbors Network ($\epsilon$-NN):} each node is connected to all the nodes whose distance is shorter than $\epsilon$, a user-defined threshold~\cite{ferreira22}.
    \item \textbf{Weighted Network:} is constructed by connecting all pairs of nodes and using their distances as weights. Typically, shorter distances correspond to stronger links, and the weighted adjacency matrix can be defined as $A = 1 - D$ or normalized $D_{norm}$~\cite{ferreira22}.
    \item \textbf{Networks with Significant Links:} connects nodes only if their distance is statistically significant~\cite{ferreira22}. For example, the significance of the Pearson correlation coefficient can be tested using the z-transformation.
\end{itemize}


\section{Data Exploration}

In this section, we introduce the fundamental datasets that underpin our study. Our primary data sources include the LOTAÇOR/OKEANOS-UAc daily landings dataset and the PNRD/OKEANOS-UAc inquiries database~\cite{LotacorSA}. These inquiries are systematically collected by samplers during fishery landings in the Azores' main fishing harbors, offering rich insights into fishing activities.

The inquiries encompass a wealth of information, including the precise locations (island and harbor) of the landings, the common names and major species groups of the captured fish, the weight of the catch, the types of fishing gear employed and other essential vessel-related details. Each individual observation within these datasets is uniquely identified and timestamped. Spanning the years from 2010 to 2017, our data comprises a total of 30,281 observations. 

\subsection{Data Description}

As there was a wide variety of different fish species, we decided to study the major fish groups (classifications) instead. To further understand each classification, it is important to consider that demersal fish live and feed on or near the bottom of water bodies~\cite{demersal}, while Pelagic fish live in the pelagic zone of ocean, which comprises the open, free waters away from the shore~\cite{pelagic}. Aditionally, there are two primary categories of demersal fish: those that are exclusively benthic and can reside on the seafloor, and those that are benthopelagic and can hover in the water column just above the seafloor~\cite{demersal}. Similarly, marine pelagic fish can be classified into two groups: pelagic coastal fish and oceanic pelagic fish~\cite{pelagic}.

Table~\ref{table:class} provides a overview of the 13 classification types utilized to categorize the fish in our data, as well as the total number of landings, the total weight of fish caught and the average weight for each classification, in kilograms (Kg).

\begin{table}[htp]
\centering
\caption{Major Fish Classifications}
\label{table:class}
\resizebox{\columnwidth}{!}{\begin{tabular}{ |p{8cm}|p{1.5cm}|p{1.5cm}|p{1.5cm}| }\hline
    \textbf{Classification} & \textbf{Landings Amount} & \textbf{Total Weight} & \textbf{Average Weight} \\\hline
    Tunas (T) & 998 & 6131140 & 115.30 \\\hline
    Continental Shelf Slope Demersals (CSSD) & 6298 & 672199 & 57.92 \\\hline
    Small Pelagics (SP) & 2886 & 505364 & 2.09 \\\hline
    Deep-Sea Species (DS) & 2028 & 474636 & 3.56 \\\hline
    Continental Shelf Slope Benthopelagic (CSSB) & 5247 & 432526 & 4.00 \\ \hline
    Demersals (D) & 3085 & 215418 & 79.10 \\\hline
    Mollusks (M) & 3181 & 104611 & 1.50 \\\hline
    Coastal Demersals (CD) & 5270 & 72392 & 7.55 \\\hline
    Large Migratory Pelagics (LMP) & 141 & 25019 & 650.0 \\\hline
    Coastal Pelagics (CP) & 877 & 23561 & 13.55\\\hline
    Small Coastal Demersals (SCD) & 158 & 1605 & 0.30 \\\hline
    Other Spp (OS) & 110 & 651 & NA\\\hline
    Crustaceans (C) & 2 & 13 & 1.50 \\\hline
\end{tabular}}
\end{table}

The calculation of average weights involved the utilization of the most frequently encountered fish species within each group, taking into account the proportions of these individual species' weights. Our focus was on those species whose combined weight contributed to 90\% of the total classification weight. We obtained the necessary weight data through the \textit{rfishbase} resource~\cite{rfishbase}.

It's also worth noting that the weight caught per fishery is influenced, not only by the average weight of the species caught, but also by the amount of fish caught in each landing, which is influenced by the behavior of different species moving together. Certain species may have a tendency to aggregate or swim together, leading to a higher catch per landing. For instance, Tunas are known to exhibit aggregative behavior, often forming schools or loose aggregations~\cite{tunas}. 

We also analysed the methods of fishing. Table~\ref{table:metier} provides a list of the 14 different types of fishing gear, also known as metiers, along with their descriptions and the corresponding number of landings and total weights associated to each.

\begin{table}[htp]
\centering
\caption{Metier Description}
\label{table:metier}
\resizebox{\columnwidth}{!}{\begin{tabular}{ |p{2cm}|p{8cm}|p{1.5cm}|p{1.5cm}| }
\hline
\textbf{Metier} & \textbf{Description} & \textbf{Landings Amount} & \textbf{Total Weight} \\ \hline
LHP-TUN & Pole-and-line for tuna species & 1002 & 6130088 \\ \hline
LLS-PD & Bottom longline & 10396 & 1203759 \\ \hline
PS-PPP & Purse seine nets for small pelagic fish & 2362 & 480957 \\ \hline
LLD-PP & Deep-water drift bottom longline & 115 & 293457 \\ \hline
LHP-PB & Bottom fish handlines & 11247 & 277624 \\ \hline
LLS-DEEP & Deep-water non-drifting bottom longline & 807 & 134733 \\ \hline
LHP-CEF & Handline jigging for catching squids & 3148 & 103761 \\ \hline
GNS-PB & Coastal gillnets & 943 & 17616 \\ \hline
LLD-GPP & Surface drifting longline for large migratory pelagics & 42 & 13595 \\ \hline
PS-PB & Lifting nets for small coastal fishes & 84 & 1375 \\ \hline
LHP-PBC & Pole-and-line and coastal trolling for small pelagics & 36 & 1071 \\ \hline
FPO-PB & Fish traps & 80 & 1014 \\ \hline
FPO-CRU & Crustacean traps & 2 & 53 \\ \hline
NEI & Not Identified & 17 & 34 \\ \hline
\end{tabular}}
\end{table}

\begin{figure}[htp]
    \centering
    \includegraphics[width=1.0\linewidth]{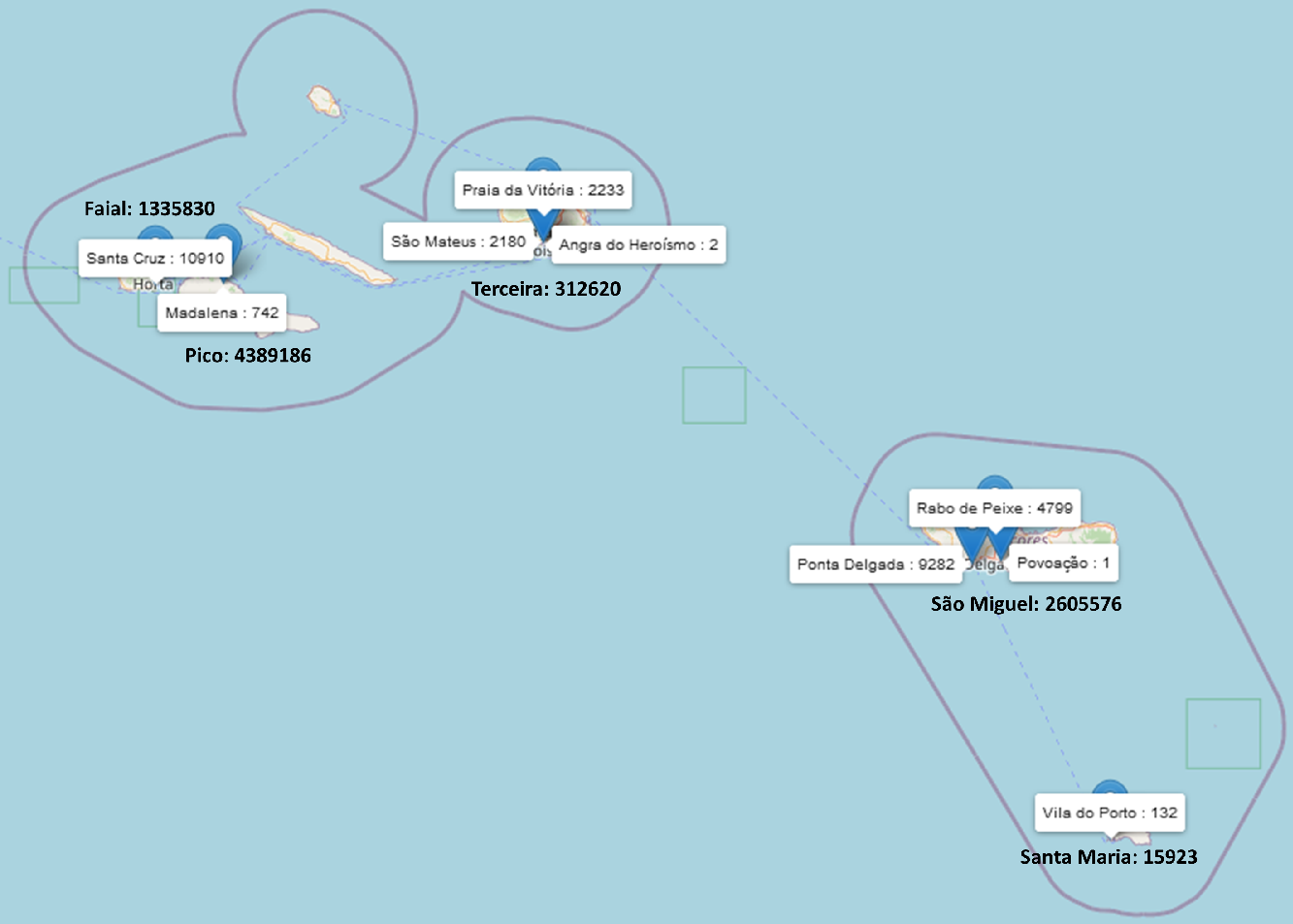}
    \caption{Azores Harbors}
    \label{fig:map}
\end{figure}

Another important factor is the location. Our data comprises 9 different harbors, that belong to 5 different islands. For visualization purposes, a map with the harbors marked is provided in Fig.~\ref{fig:map}, with the total weight of fishes caught associated to each of the islands (labels without background) and the number of landings registered for each harbor (labels with white background). 

\subsection{Data Preparation}

We observed a data gap from January 2014 to March 2014 and opted to fill it using data from the corresponding period in 2013, as it exhibited a similar number of landings. However, given the substantial differences in weight values between 2013 and 2014, we introduced a scaling factor. This was determined by comparing total weights from April to December in both years. We then applied this factor to the 2013 data to estimate the missing weight values for 2014.

We intentionally excluded data from the harbors Angra do Heroísmo and Povoação, the classification Crustaceans, and the metier FPO-CRU due to limited data availability and low landing volumes. Additionally, the classification Other Spp and the metier NEI were also omitted from our analysis as they had limited relevance. These exclusions were made to streamline our network analysis and align it more closely with our research objectives.

\begin{figure}[htp]
\centering
\centerline{\includegraphics[width=1.0\linewidth,height=8cm]{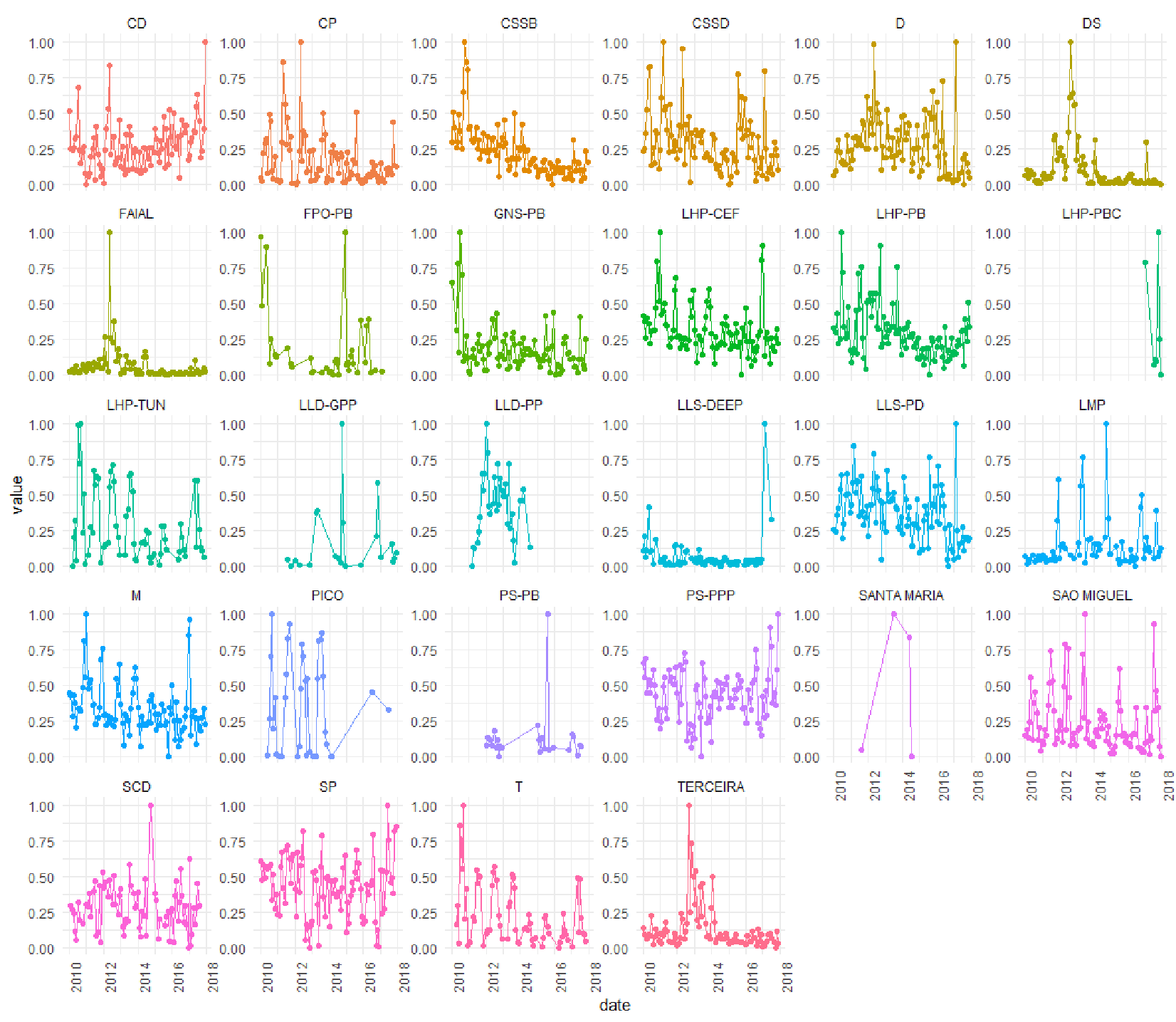}}
\caption{Time Series of Islands, Metiers and Classifications}
\label{fig:timeseries}
\end{figure}

After completing the imputation and data cleaning steps, we proceeded to generate the time series. To construct these, we aggregated the observations by calculating the mean value for each classification, metier, and island, per month, and then normalized each series, as presented in Figure~\ref{fig:timeseries}. Notably, we excluded harbors from this aggregation since they are closely tied to their respective islands and could introduce more complexity than desired.

\section{Experimental Analysis}

Understanding the dynamic changes in fisheries practices over the years is a challenging and complex task. We delved into three key questions related to alterations in the structure of networks:

\begin{enumerate}
    \item In transforming distance values to a network, which method yields a more suitable network for investigating dynamic changes?
    \item Which connections undergo changes over time, and which ones exhibit consistent patterns in the realms of classification-metier, classification-island, and classification-classification connections?
    \item Which nodes demonstrate higher interaction levels with others in each year?
\end{enumerate}

Initially, our objective was to identify the most suitable methods for transforming the distance matrix into an adjacency matrix for a dynamic study. Given our emphasis on observing variations in community structure over the years, we examined modularity values for each year. Additionally, to avoid an excessively sparse network with numerous isolated nodes, we investigated the network density. In conclusion, we observed the trade-off between these characteristics.

Next, we examined changes in networks, focusing on connections between classifications and fishery gears over the years. This allowed us to identify classifications with limited flexibility regarding fishery gears and those that were more adaptable. Additionally, we explored connections between classifications and islands to discern any migratory patterns or changes in species habitats over time. Finally, we observed connections between classifications to identify consistent patterns of fishery across species.

Finally, we investigated the nodes with the highest degree in each year, reflecting those with stronger connections to other nodes. We observed whether this pattern remained consistent over the years or underwent changes. 

In conclusion, our study aimed to examine intrinsic changes in the dynamics of fisheries through a visible and flexible method.

\subsection{Methods}

We decided to create multiple networks, one for each year from 2010 to 2017. This decision aligns with our goal of uncovering relationships among different factors, showing how they influence or remain unaffected by others, over time. In this case, each network will have as node either an island, a metier or a fish group and the edges between nodes signify a strong similarity.

To transform our set of time series into networks, we employed the `ts2net`~\cite{ferreira22} library in R, with Dynamic Time Warping (DTW) being the distance function used, as justified in~\ref{literature}. 

\subsection{Results}

This section provides an overview of the answers to the research questions proposed in this study. 

\subsubsection{Network Construction:}In our exploration of different network construction approaches, we tested the methods described in Section~\ref{literature}: $k-NN$ network, $\epsilon-NN$ network, weighted network and significant links network. We considered various parameter values, specifically, different values for $k$ (2, 3, 5, 7, and 10) and $\epsilon$ (0.3, 0.5, 0.7, and 0.9). 

Although the network with significant links initially exhibited the highest mean modularity, we ultimately selected the k-NN network with $k$ =$2$ as our preferred choice. This decision was made because the network with significant links suffered from a lack of connections, resulting in a highly fragmented and disconnected network with a mean density of 0.02656478, while $k$ =$2$ shows a mean density of 0.11386721, which is sparse network, but not a fragmented one, being more suitable for the goals of our study.

\subsubsection{Network Analysis:} 
\begin{figure}[htp]
    \centering
    \includegraphics[width=1.0\linewidth,height=7cm]{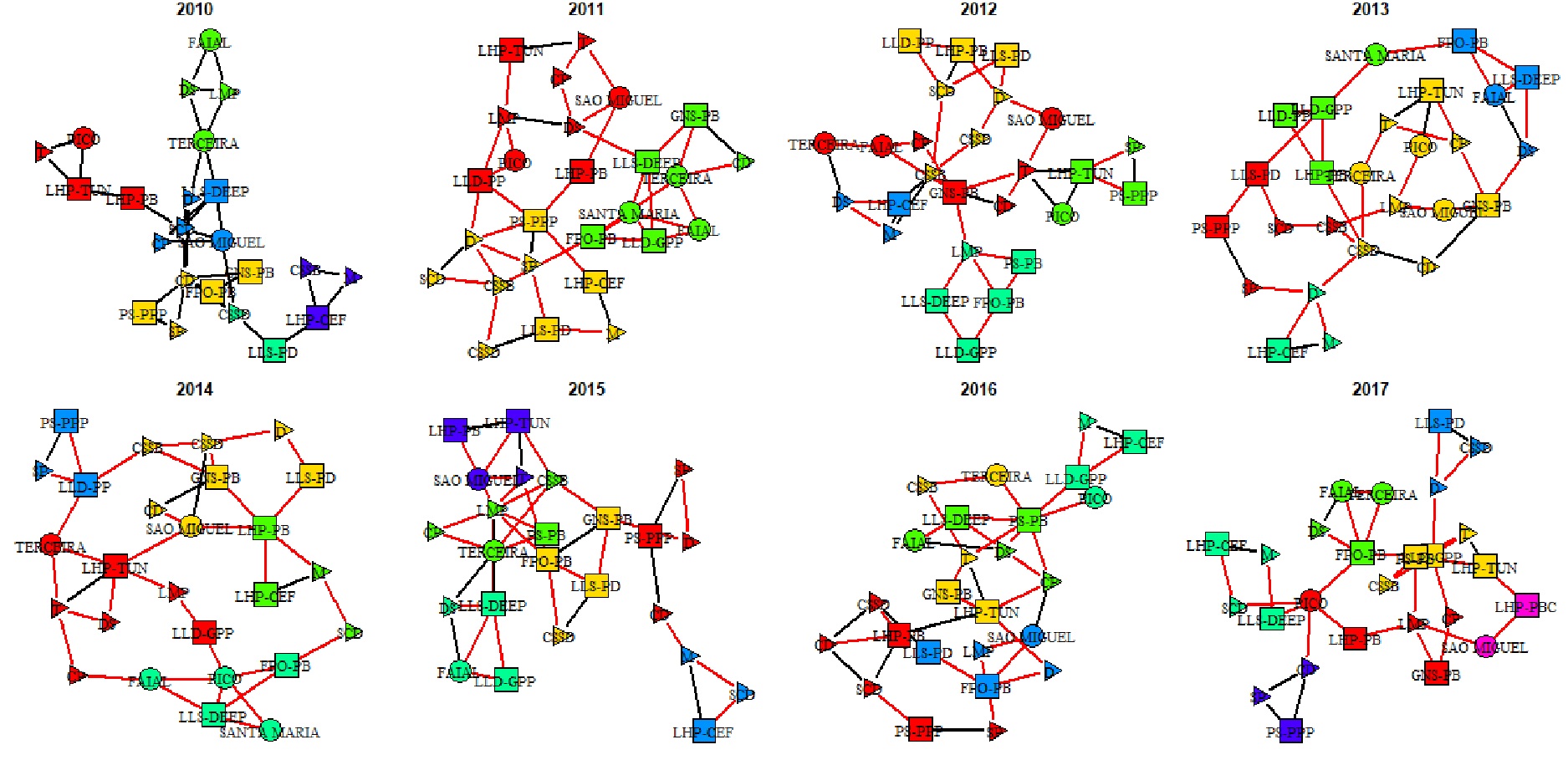}
    \caption{Networks over the years}
    \label{fig:networks}
\end{figure}

For visualization purposes, we employed the `igraph` package. In Fig.~\ref{fig:networks}, the $2$-NN networks for each year are depicted. Red edges signify new connections formed from one year to the next, while black edges represent retained connections. Triangle-shaped nodes correspond to classifications, circle-shaped nodes represent islands, and square-shaped nodes denote metiers to a better visualization. Nodes are color-coded to signify their communities in each year, identified using the "cluster\_walktrap" function from the `igraph` package~\cite{igraph}. This function identifies communities based on random walks. 

Clearly, the rate of new edges that emerges between years is significantly high, demonstrating a great changing in the network over the years. 

\paragraph{class-metier} Notably, certain associations remain steady over the years, such as the evident links between Tunas and LHP-TUN, Mollusks and LHP-CEF, Small Pelagics and PS-PPP, as well as Deep-sea Species and LSS-DEEP. More specific interactions involve CSS Demersals and LLS-PD, which are prominently associated in the initial years, followed by a gap in utilization, but then a resurgence. Linked to Coastal Demersals, we observe GNS-PB, until 2014, and PS-PPP, particularly more recently.

Conversely, several classifications, including Continental Shelf Slope (CSS) Benthopelagics, Demersals, and Large Migratory Pelagics, demonstrate a diverse use of or no connection to fishing gears. This lack of consistent patterns suggests a flexible approach to fishing methods. Small Coastal Demersal present varying methods employed, transitioning from LHP-BP initially to LHP-CEF more recently, highlighting adaptability in fishing practices.

\paragraph{class-islands} Examining island connections reveals intriguing trends. Tunas exhibit strong links with São Miguel, Pico, Terceira, and other islands in various years. However, these connections seem to have diminished in recent years. As the connection with Tunas decreases, Large Migratory Pelagics appear to play a more pivotal role in São Miguel's fishing activities.

Faial displays connections to Deep-sea Species in recent years, contrasting with its past association with Coastal Pelagics. Meanwhile, Terceira exhibits connections to Deep-sea Species and CSS Benthopelagics, although only during specific years, indicating variability in fishing patterns. Santa Maria stands out for its lack of strong connections to classifications, reflecting a diverse and potentially evolving ecosystem within the region.


\paragraph{class-class} Furthermore, examining links between classifications themselves, reveals intriguing dynamics. Robust connections persist between Demersal-related categories and Small Pelagics over the years. On the opposite hand, the associations between Tunas and Coastal Pelagics seem to have diminished recently. 

Furthermore, Deep-sea Species and Large Migratory Pelagics displayed a strong connection in earlier years, but this link has since waned. As for Mollusks, similar patterns to both CSS Benthopelagics and Small Coastal Demersals are observed, but during different time periods, highlighting temporal variations in their connections.

\subsubsection{High Degree nodes:} Nodes with high degrees often indicate the ecological importance of a species, the adaptability of a fishing gear, or the significance of an island within the fishing network for a specific year. Key nodes in the analysis include FPO-PB, maintaining a consistently high degree in the last three years: a degree of 5 in 2015 and 2016, and 6 in 2017. PS-PB also stands out, demonstrating significance in the last two years with a degree of 7 in 2016 and 6 in 2017. GNS-PB emerges as important in 2012 with a degree of 6 and in 2013 with a degree of 4, while LHP-TUN appears in 2014 and 2016, both with a degree of 5, alongside LLD-GPP in 2013 with a degree of 4 and 2017 with a degree of 6.

\subsection{Discussion}
Our study sheds light on a dynamic and evolving fishing network, where relationships among fish classifications, fishing methods (metiers), and geographical locations (islands) constantly shift.

Tunas emerge as a prominent contributor to the overall weight of fisheries in the Azores. However, as time has progressed, we've discerned a concerning decline in the total catch weight of these species. This decline, coupled with diminishing connections to once-consistent islands, raises pressing questions. It compels us to consider the sustainability of tuna populations and the possibility of shifts in their migration patterns. This underscores the paramount need for vigilant fisheries management in these regions.

Moreover, the vanishing connections involving various fish classifications and Faial in more recent years raise another layer of concern. They suggest a noteworthy alteration in the marine ecosystem around the island over the years.

Within our network analysis, specific nodes also stand out. Notably, FPO-PB and PS-PB emerged as pivotal nodes recently, highlighting the importance of specific fishing methods. 


\section{Conclusions}
In this paper, we offer a complex network analysis approach, that has offered profound insights into the temporal trends detected within our time series data. By unveiling intricate relationships across various features and identifying critical nodes within the network, we've not only shed light on the changes observed over time but also acquired a more profound understanding of the complex dynamics of Azorean fisheries.

This understanding isn't just informative, as network analysis emerges as a pivotal tool in real-world scenarios. Particularly, in identifying challenges in fisheries management, it can aid the critical decision-making processes, particularly concerning quota definition and tracking, ensuring the sustainability of marine ecosystems and the livelihoods of those dependent on them.

Our analysis represents just one aspect of fisheries research, emphasizing the ongoing need for further investigations and collaborations in this critical field. Additionally, the flexibility of `ts2net`, with its ability to explore various parameter settings, offers diverse insights into the dynamics of Azores fisheries. As a potential avenue for future research, we could explore other approaches like NetF~\cite{Silva2022}, which transforms a single time series into a network using quantile and visibility graphs, extracting significant topological measures. This could provide valuable additional perspectives on the subject. 

In conclusion, our study reaffirms the importance of complex network analysis in real word data. By translating intricate time series into networks and exploring their properties, we gain valuable insights into temporal fisheries dynamics, essential to guiding us toward sustainable practices and emphasizing the urgency of continued research and data collection in this intricate and ever-changing marine ecosystem.

%
%
\bibliographystyle{spmpsci} 
\bibliography{main} 
\end{document}